# Thickness dependence of spin Hall magnetoresistance in FeMn/Pt bilayers

Yumeng Yang[1,2], Yanjun Xu[1], Kui Yao[2], and Yihong Wu[1,a)]

[1]*Department of Electrical and Computer Engineering, National University of Singapore, 4 Engineering Drive 3, Singapore 117583, Singapore*

[2]*Institute of Materials Research and Engineering, A*STAR (Agency for Science, Technology and Research), 2 Fusionopolis Way, 08-03 Innovis, Singapore 138634, Singapore*

We investigated spin Hall magnetoresistance in FeMn/Pt bilayers, which was found to be one order of magnitude larger than that of heavy metal and insulating ferromagnet or antiferromagnet bilayer systems, and comparable to that of NiFe/Pt bilayers. The spin Hall magnetoresistance shows a non-monotonic dependence on the thicknesses of both FeMn and Pt. The former can be accounted for by the thickness dependence of net magnetization in FeMn thin films, whereas the latter is mainly due to spin accumulation and diffusion in Pt. Through analysis of the Pt thickness dependence, the spin Hall angle, spin diffusion length of Pt and the real part of spin mixing conductance were determined to be 0.2, 1.1 nm, and $5.5 \times 10^{14}\,\Omega^{-1}\,m^{-2}$, respectively. The results corroborate the spin orbit torque effect observed in this system recently.

a) Author to whom correspondence should be addressed: elewuyh@nus.edu.sg

Recently unconventional magnetoresistance (MR) has been reported in a variety of ferromagnet (FM) / heavy metal (HM) bilayers, with the FMs including both ferromagnetic insulators such as yttrium iron garnet (YIG),[1-12] $CoFe_2O_4$,[13] $NiFe_2O_4$,[4] $Fe_3O_4$[4,14] and $LaCoO_3$[15] and ferromagnetic metals, *e.g.*, Co,[16,17] CoFeB,[18-20] and NiFe,[21] and the HMs including Pt,[4-8,10,13-15] Pd,[9] Ta,[5,11] and Ru.[22] Despite its debatable origin,[2,3,23-25] the experimental data reported to date seem to support the spin Hall magnetoresistance (SMR) theory.[3-14,16-20,22] In the SMR scenario, the charge current passing a thin HM layer generates a transverse spin current in the thickness direction via the spin Hall effect (SHE). The spin current is partially reflected back to the HM layer when it reaches the FM/HM interface, with the reflection coefficient determined by the angle between the polarization of spin current ($\vec{\sigma}$) and magnetization direction of the FM ($\vec{m}$). The reflected spin current in turn generates an additional charge current through the inverse spin Hall effect (ISHE), leading to the appearance of SMR: $R_{xx} = R_0 - \Delta R(\vec{m} \cdot \vec{\sigma})^2$, where $R_{xx}$ is the longitudinal resistance, $R_0$ the isotropic longitudinal resistance, and $\Delta R$ the change in resistance induced by the SMR effect. In addition to FMs, the SMR has also been observed in $SrMnO_3$/Pt[26] in which $SrMnO_3$ is an antiferromagnetic (AFM) insulator. Compared to the FM/HM bilayers, the MR behavior of AFM/HM is more complex as the spin state of AFM is strongly dependent on its thickness. When viewed from a different perspective, however, this sensitivity to thickness provides a convenient way to study how the SMR is related to both the magnitude and direction of AFM magnetization through varying its thickness systematically. The results obtained shall shed light on spin orbit torque (SOT) in AFM/HM bilayers, a phenomenon that is closely related to the SMR.

In view of the above, we investigated both the Pt and FeMn thickness dependences of SMR in FeMn/Pt bilayers, a system where we have recently observed a large SOT effect due to the presence of small net magnetic moments in FeMn.[27] Through angle-dependent magnetoresistance (ADMR) measurement, it was found that the MR observed in FeMn/Pt is dominantly from the SMR origin. The size of the MR is on the order of $10^{-3}$, one order of magnitude larger than that of SrMnO$_3$/Pt bilayers,[26] and is comparable to that of NiFe/Pt bilayers. The results are in good agreement with the strong SOT effect observed in FeMn/Pt bilayers. A clear FeMn thickness dependence of SMR was observed, which can be understood by taking into account the spin transport in both FeMn and Pt layers, and thickness dependence of net magnetization in FeMn. It is found that the latter plays a more dominant role in determining the FeMn thickness dependence of SMR in FeMn/Pt bilayers, in a sharp contrast to NiFe/Pt bilayers wherein the NiFe thickness effect is mainly attributed to the spin transport in both NiFe and Pt layers above a certain NiFe thickness. In addition, through Pt thickness dependence analysis of SMR in FeMn/Pt, the spin Hall angle ($\theta_{SH}$), spin diffusion length ($\lambda_{Pt}$) of Pt and the real part of spin mixing conductance (Re[$G_{MIX}$]) were determined to be 0.2, 1.1 nm, and $5.5 \times 10^{14}\,\Omega^{-1}\,m^{-2}$, respectively.

Two series of FeMn/Pt bilayer samples in the form of Hall bars were prepared on SiO$_2$ (300 nm)/Si substrates (starting from the FeMn layer) using combined techniques of sputtering (base pressure: $3\times10^{-8}$ Torr and process pressure: 3 mTorr) and photolithography. The two series of samples have the structure of (i) FeMn($t_{FeMn}$)/Pt(3) and (ii) FeMn(3)/Pt($t_{Pt}$) (number inside the parentheses indicates the thickness in nm). The thicknesses of FeMn ($t_{FeMn}$) and Pt ($t_{Pt}$) were varied in the range of 0.5 – 15 nm and 1 – 15 nm, respectively. In addition, we also fabricated a series of NiFe($t_{NiFe}$)/Pt(3) control samples with $t_{NiFe}$ = 0.5 – 15 nm for comparison purpose. The thickness of each layer is precisely controlled based on pre-calibrated deposition rates, which were 0.046 nm/s, 0.055 nm/s and 0.075 nm/s for FeMn, NiFe and Pt, respectively. The relatively low deposition rates allow for precise thickness control by adjusting the deposition duration of each layer. The Hall bars have a dimension of 2.3 mm × 0.2 mm in the central region and 0.1 mm × 1 mm for the transverse electrodes. The resistivity of individual layers was extracted from the overall resistivity of FeMn/Pt bilayers with different thickness combinations based on a parallel-resistor model, and the obtained resistivity values are: $\rho_{FeMn}$ = 166 μΩ·cm, $\rho_{NiFe}$ = 40 μΩ·cm, and $\rho_{Pt}$ = 32 μΩ·cm, respectively.

In the first round of measurements, conventional field-dependent MR measurements were performed on series (i) samples with $t_{FeMn}$ = 0.5 – 15 nm with the sweeping magnetic field $H$ applied in $x$-, $y$-, $z$- direction, respectively. Figs. 1(a) – 1(c) show the field-dependent MR results for samples with $t_{FeMn}$ = 0.5 nm, 3 nm, and 15 nm, respectively. The general observations are (i) the MRs in $x$- and $z$-directions have the same polarity and are much larger than the MR in $y$-direction, and (ii) the MR in $z$-direction cannot be explained by the conventional anisotropic MR behavior,[3,11] which should give a negative MR in z-direction when the current is applied in x-direction. From exchange bias studies of FeMn/NiFe bilayers,[27] it was found that FeMn starts to show the onset of clear exchange bias only at a thickness around 4 - 5 nm. At $t_{FeMn}$ = 0.5 nm, the FeMn can be considered as a superpara-antiferromagnet at room temperature when it is standalone; however, when contacted with Pt, it behaves more like a FM due to interaction with Pt. When $t_{FeMn}$ increases to 3 nm, weak AFM order appears as reflected in the enhancement of coercivity in FeMn/NiFe bilayers. Therefore, in both the $t_{FeMn}$ = 0.5 nm and 3 nm samples, there is significant amount of uncompensated spins in the FeMn layer and their spin sub-lattices can be rotated easily by the external field. When $t_{FeMn}$ increases further to 15 nm, the AFM order becomes more rigid and difficult to be rotated by the external field. In this case, it is the uncompensated spins at the interface that are responsible for the MR observed. On the other hand, the MR behavior of FeMn/Pt is found to be insensitive to the change in Pt thickness at a fixed FeMn thickness. Based on these considerations and the strong dependence of both the magnitude and curve shape of MR on $t_{FeMn}$, it is apparent that the MR observed in the FeMn/Pt samples is closely related to the spin configuration of FeMn. The same polarity of MR in $x$- and $z$-directions suggests that the MR observed is of SMR origin. Although the so-called Hanle effect MR induced in the Pt layer itself also has the same polarity,[24] its size on the order of $10^{-6}$, as verified by a Pt(3)/SiO$_2$/Si control sample, is too small to account for the MR observed in FeMn/Pt bilayers.

In order to extract the SMR contribution from the overall MR, ADMR measurements were performed on these bilayers. As illustrated in Fig. 2(a), the longitudinal resistance of the sample was measured while rotating a constant field $H$ in $zy$, $zx$, and $xy$ planes, respectively. The SMR ratio is calculated from the relation $\Delta R / R_{xx} = (R_{xx}^z - R_{xx}^y) / R_{xx}^y$, where $R_{xx}^z$ and $R_{xx}^y$ are the longitudinal resistance when the magnetization is saturated in $z$- and $y$- direction, respectively. Fig. 2(b) shows the ADMR curves for FeMn(3)/Pt(3) measured with a constant field of 30 kOe, which are representative of FeMn/Pt bilayers with different thickness combinations. From Fig. 1(b), it can be seen that 30 kOe is large enough to saturate the magnetization of the bilayers in the field direction. The general features of the ADMR curves can be summarized as follows: (i) the $\theta_{zx}$-dependence of MR is vanishingly small; (ii) $\theta_{zy}$- and $\theta_{xy}$-dependences of MR are much stronger than that on $\theta_{zx}$ and the two curves almost overlap with each other. The vanishing $\theta_{zx}$-dependence of MR indicates that the conventional anisotropic magnetoresistance (AMR) is negligibly small in FeMn/Pt bilayers (note that the SMR should be zero in this measurement configuration). This in combination with the almost overlapping $\theta_{zy}$- and $\theta_{xy}$-dependence of MR again demonstrates clearly that the MR in FeMn/Pt is dominated by SMR. The SMR ratio, on the order of $10^{-3}$, is one order of magnitude larger than that of the SrMnO$_3$/Pt system.[26] Fig. 2(c) shows the ADMR results of a NiFe(3)/Pt(3) bilayer measured in the same configurations for comparison. The main difference with FeMn(3)/Pt(3) is that the AMR ($\theta_{zx}$-dependence of MR) is much larger in this case, which causes a clear separation between the $\theta_{xy}$- and $\theta_{zy}$-dependence of MR curves. It is apparent that the sum of MR measured with the field rotating in the $zx$ and $zy$ plane is equal to that measured when the field rotates in the $xy$ plane. It is also worth noting that the magnitudes of SMR in both systems are similar.

To have a more quantitative understanding of the SMR effect in FeMn/Pt bilayers, we investigated the thickness dependence of the effect for each layer. Fig. 3(a) shows the $\theta_{zy}$-dependence of MR for the FeMn(3)/Pt($t_{Pt}$) series of samples with $t_{Pt}$ = 1 nm, 2 nm, 5 nm, 8 nm and 15 nm, respectively. As summarized in Fig. 3(b), the SMR ratio shows a non-monotonic dependence on the Pt thickness; it increases initially at small thicknesses, peaks at about 3 nm, and then decreases between 3 – 15 nm. The $t_{Pt}$-dependence of SMR is similar to those observed in CoFeB-based FM/HM bilayers.[18-20] When dealing with metallic FM/HM bilayers, one has to take into account both the charge current shunting effect[18] and the longitudinal spin current that travels into the metallic FM layer.[19,28] Following the drift-diffusion formalism, Kim *et al.* have derived an expression for SMR in FM/HM bilayers:[19]

$$\frac{\Delta R}{R_{xx}} = \theta_{SH}^2 \frac{\lambda_N}{d_N} \frac{\tanh^2(d_N/2\lambda_N)}{1+\xi} \left[ \frac{g_r}{1+g_r \coth(d_N/2\lambda_N)} - \frac{g_F}{1+g_F \coth(d_N/2\lambda_N)} \right] \quad (1)$$

with $g_r = \rho_N \lambda_N \text{Re}[G_{MIX}]$, and $g_F = \frac{(1-P_C^2)\rho_N \lambda_N}{\rho_F \lambda_F \coth(d_F/\lambda_F)}$. Here, $\rho_N$, $\rho_F$, $\lambda_N$, $\lambda_F$, and $d_N$, $d_F$ are the resistivity, spin diffusion length and thickness of HM and FM, respectively, $\text{Re}[G_{MIX}]$ is the real part of the spin mixing conductance, $P_C$ is the current spin polarization of FM, and $\xi = \rho_N d_F / \rho_F d_N$ is the current shunting factor. For the case of insulating FM/HM system, since $\xi = 0$ and $g_F = 0$ ($\rho_F \to \infty$), the SMR is only determined by the first term inside the square bracket of Eq. (1).[23] The second term is included to account for the longitudinal spin current traveling inside the FM driven by the spin accumulation at the FM/HM interface. Compared with the FM insulator case, the largest correction of SMR happens when $P_C$ approaches 0. In this case, the FM layer is essentially a non-magnetic metal (NM); therefore, the SMR diminishes except for the very small Hanle MR.[24] On the other hand, when $P_C$ approaches unity, the FM becomes a half-metal. In this case, the spin current cannot flow vertically in the FM layer due to lack of minority spin carriers and thus there will be no additional correction to SMR except for the current shunting effect. The situation is more complex in FeMn/Pt bilayers, in particular when FeMn is thin. In this case, the FeMn is neither a good AFM nor an FM; its spin structure depends strongly on the thickness. Considering the much smaller spin Hall angle[29,30] and larger resistivity of FeMn as compared to Pt, the spin current generated in FeMn can be neglected. The SMR of FeMn/Pt bilayers is dominantly due to the spin current in Pt. Therefore, without losing generality, we may still use Eq. (1) to model the SMR dependence on FeMn thickness, but we have to introduce a thickness-dependent polarization for FeMn. This is a reasonable approach because when $t_{FeMn}$ is large, a rigid AFM order will form which results in diminishing polarization. On the other hand, when $t_{FeMn}$ is small (*e.g.*, $t_{FeMn}$ = 3 nm), the net magnetic moment induced by an external field shall lead to a non-zero $P_C$ value. Based on these considerations, we first analyze the $t_{Pt}$-dependence of SMR with a constant $P_C$ value and then discuss the $t_{FeMn}$-dependence by taking into account the thickness dependence of polarization, which can be inferred from the magnetization data.

As shown in Fig. 3(b), the $t_{Pt}$-dependence of SMR can be fitted reasonably well using Eq. (1) (solid line) with fitting parameters: $P_C$ = 0.37, $\theta_{SH}$ = 0.2, $\lambda_{Pt}$ = 1.1 nm, $\lambda_{FeMn}$ = 2.0 nm and Re[$G_{MIX}$] = 5.5 × $10^{14}$ $\Omega^{-1}$ m$^{-2}$. Note that the $P_C$ value used here is obtained from the $t_{FeMn}$-dependence of magnetization which will be discussed shortly. It should also be noted that it is not possible to obtain $\theta_{SH}$ and Re[$G_{MIX}$] independently based only on SMR results since the value used for one would affect the other. Therefore, during the fitting, we set $\theta_{SH}$ = 0.2 and treat Re[$G_{MIX}$] as a fitting parameter. This is a reasonable assumption considering the fact that the intrinsic Hall angle for Pt is reported to be in the range of 0.15 - 0.3.[31-33] As can be seen from the figure, the fitting agrees quite well with the experiment data. The fitting values for $\lambda_{Pt}$ and $\lambda_{FeMn}$ are comparable with the reported values for Pt[4,6,10] and FeMn,[27,29,34] respectively. The results indicate that the drift-diffusion model can satisfactorily describe the spin current generation and transport in FeMn/Pt bilayers at a fixed FeMn thickness.

We now turn to the $t_{FeMn}$-dependence of SMR in the bilayers. Fig. 4(a) shows the $\theta_{zy}$-dependence of MR for FeMn($t_{FeMn}$)/Pt(3) bilayers with $t_{FeMn}$ = 0.5 nm, 2 nm, 5 nm, 8 nm, and 15 nm, respectively. For comparison, we also show in Fig. 4(b) the $\theta_{zy}$-dependence of MR for NiFe($t_{NiFe}$)/Pt(3) control samples with $t_{NiFe}$ = 0.5 nm, 2 nm, 5 nm, 8 nm, and 15 nm, respectively. Both series of samples exhibit clear SMR behavior with its magnitude depending on the FeMn or NiFe thickness. It is worth noting that the maximum SMR ratios of the two series of samples are almost the same (2.54 × 10$^{-3}$ for FeMn/Pt and 2.49 × 10$^{-3}$ for NiFe/Pt). The detailed $t_{NiFe}$ and $t_{FeMn}$ dependences of SMR are shown in Fig. 4(c) and Fig. 4(d), respectively. Similar to the Pt thickness dependence shown in Fig. 3 (b), a non-monotonic dependence on $t_{FeMn}$ or $t_{NiFe}$ is obtained. Despite the fact that the $t_{Pt}$-dependence of SMR can be explained reasonably well using Eq. (1), the same equation is unable to fit the $t_{FeMn}$-dependence if we use a fixed $P_C$ value. As mentioned above, to account for $t_{FeMn}$-dependence, it is necessary to use a $t_{FeMn}$-dependent $P_C$ value for FeMn. It is noticed that in metallic FMs, the tunneling spin polarization ($P_T$) is approximately linear to the magnetization, *i.e.*, $P_T \propto M_s$.[35,36] As a first approximation, we assume that the same relation also holds for current spin polarization ($P_C$) used in Eq. (1) and net magnetization in thin AFM layers. This is supported by the fact that: i) $P_C$ determined by point-contact Andreev reflection spectroscopy is similar to $P_T$ determined by the superconductor tunneling spectroscopy for many transition metallic FMs;[37] ii) sizable net moment can be induced in FeMn by an external field (30 kOe in the SMR measurement). In this sense, we may correlate $P_C$ with the net magnetization $M_s$ of FeMn obtained by magnetometry measurements. Fig. 4(e) shows the thickness dependence of $M_s$ for FeMn at $H$ = 30 kOe extracted from the M-H loops of coupon films with the same thickness combination as the Hall bar samples.[27] As can be seen, the non-monotonic $t_{FeMn}$-dependence of $M_s$ resembles that of SMR (Fig. 4(c)) with a peak at round 2 nm, which suggests that the $t_{FeMn}$-dependence of SMR is closely related to the spin structure of FeMn. More quantitatively, we introduce a phenomenological expression for the current spin polarization $P_C(t_{FeMn}) = \alpha M_s(t_{FeMn})$, where $\alpha$ is a fitting parameter and $M_s(t_{FeMn})$ is the measured magnetization at different thicknesses. Using the parameters ($\theta_{SH}$, Re[$G_{MIX}$], $\lambda_{Pt}$, $\lambda_{FeMn}$) obtained from the fitting of Fig. 3(b) and $M_s(t_{FeMn})$ in Fig. 4(e), the $t_{FeMn}$-dependence in Fig. 4(c) can be reproduced well (solid line) with a constant $\alpha$ value of $3.1 \times 10^{-3}$ emu$^{-1}$ cm$^3$, especially at $t_{FeMn} > 2$ nm. The deviation at small $t_{FeMn}$ below 2 nm may be caused by the roughness and surface effect. It is worth emphasizing again that the curve cannot be fitted at all if we use a constant $P_C$ value. This suggests that the $t_{FeMn}$-dependence of SMR is mainly determined by the $t_{FeMn}$-dependence of net magnetization in FeMn induced by an external field. Similar thickness dependence has also been observed in the investigation of spin orbit torque effective field in FeMn/Pt bilayers.[27] On the contrary, the $t_{NiFe}$-dependence of SMR at large $t_{NiFe}$ can be well reproduced (solid line in Fig. 4(d)) using Eq. (1) with $P_C = 0.4$, $\theta_{SH} = 0.2$, $\lambda_{Pt} = 1.1$ nm, $\lambda_{NiFe} = 4.0$ nm and Re[$G_{MIX}$] = $1.2 \times 10^{15}$ $\Omega^{-1}$ cm$^{-2}$. The $P_C$ and $\lambda_{NiFe}$ values are from literature.[38,39] The deviation at $t_{NiFe} < 3$ nm can also be attributed to the roughness and surface effect which has not been taken into account in Eq. (1). These results imply that the SMR is not just an interface effect. The presence of magnetic moment in the layer adjacent to the heavy metal is crucial to obtain a large SMR. It also explains why the SMR is closely related to the spin orbit torque effect in FM/HM and AFM/HM bilayers.

In conclusion, we observed a large SMR in FeMn/Pt bilayers, with a magnitude comparable to that of NiFe/Pt bilayers. A clear FeMn thickness dependence of SMR is observed, which is mainly attributed to the thickness dependence of the net magnetization in FeMn induced by an external field. This is different from the NiFe/Pt bilayers in which the NiFe thickness dependence of SMR is mainly caused by the spin transport in both layers. Our findings shed light on interactions of spin current with spin sub-lattices in AFMs and corroborate the spin orbit torque effect observed in this system recently.

**ACKNOWLEDGMENTS**

The authors wish to thank Dr. Xiaowei Wang for his help in the data fitting. Y.H.W. would like to acknowledge support by the Singapore National Research Foundation, Prime Minister's Office, under its Competitive Research Programme (Grant No. NRF-CRP10-2012-03) and Ministry of Education, Singapore under its Tier 2 Grant (Grant No. MOE2013-T2-2-096). Y.H.W. is a member of the Singapore Spintronics Consortium (SG-SPIN).

**FIGURE CAPTIONS**

FIG. 1. (a) Field-dependent MR for FeMn($t_{FeMn}$)/Pt bilayers with (a) $t_{FeMn}$ = 0.5 nm; (b) $t_{FeMn}$ = 3 nm; (c) $t_{FeMn}$ = 15 nm.

FIG. 2. (a) Schematics for ADMR measurement with the applied field rotating in *zy*, *zx*, and *xy* planes, respectively; (b) ADMR results for FeMn(3)/Pt(3) bilayer; (c) ADMR results for NiFe(3)/Pt(3) bilayer. The results of both (b) and (c) are obtained with an applied field of 30 kOe.

FIG. 3. (a) $\theta_{zy}$-dependence of MR for FeMn(3)/Pt($t_{Pt}$) bilayers with $t_{Pt}$ = 1 nm, 2 nm, 5 nm, 8 nm and 15 nm, respectively; (b) Pt thickness dependence of SMR ratio for FeMn(3)/Pt($t_{Pt}$) bilayers (open circle: experimental data, solid-line: fitting results using Eq. (1)).

FIG. 4. (a) $\theta_{zy}$-dependence of MR for FeMn($t_{FeMn}$)/Pt(3) bilayers with $t_{FeMn}$ = 0.5 nm, 2 nm, 5 nm, 8 nm and 15 nm, respectively; (b) $\theta_{zy}$-dependence of MR for NiFe($t_{NiFe}$)/Pt(3) bilayers with $t_{NiFe}$ = 0.5 nm, 2 nm, 5 nm, 8 nm and 15 nm, respectively; (c) FeMn thickness dependence of SMR ratio for FeMn($t_{FeMn}$)/Pt(3) bilayers (open square: experimental data, solid-line: fitting results using Eq. (1)); (d) NiFe thickness dependence of SMR ratio for NiFe($t_{NiFe}$)/Pt(3) bilayers (open diamond: experimental data, solid-line: fitting results using Eq. (1)); (e) $M_s$ at $H$ = 30 kOe as a function of FeMn thickness [Ref. 27].

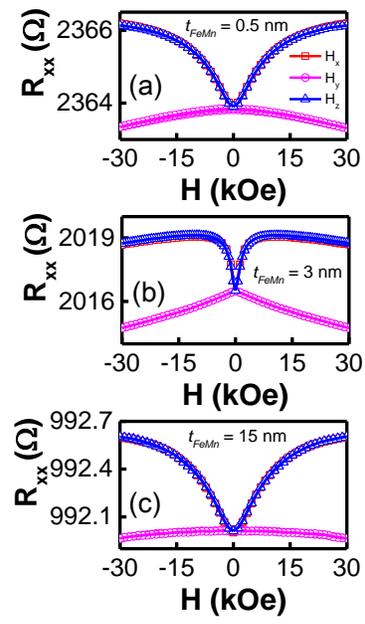

FIG. 1

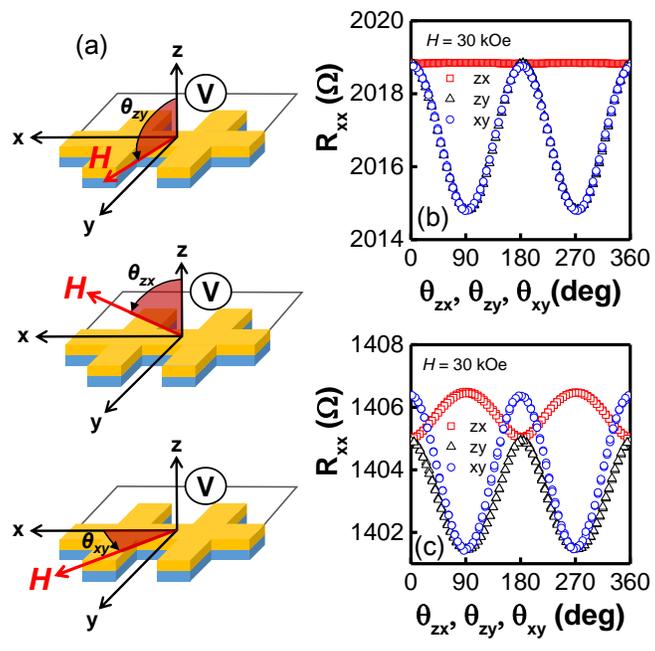

FIG. 2

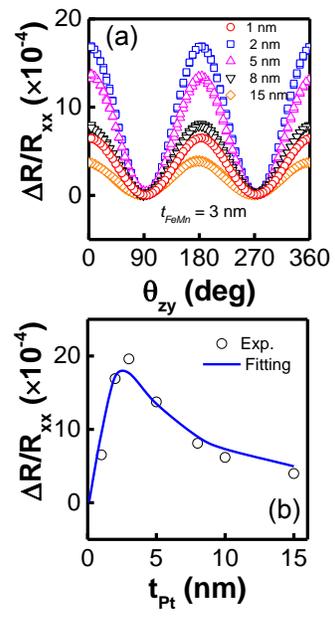

FIG. 3

Yumeng Yang

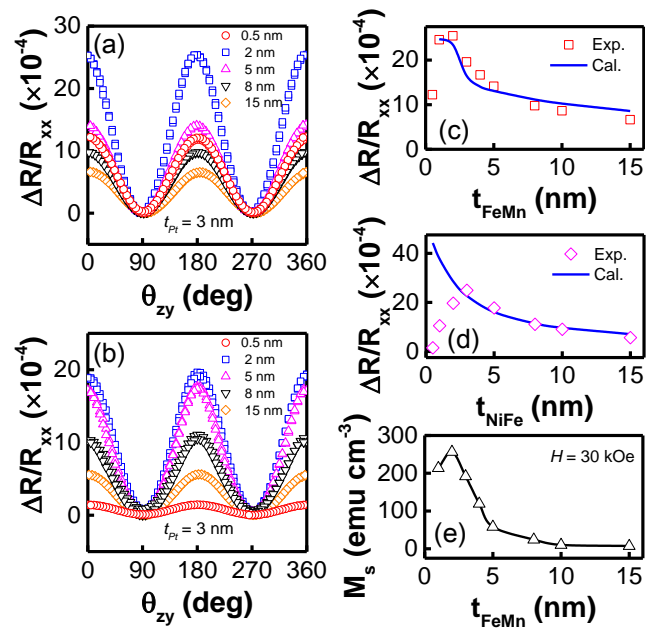

FIG. 4

Applied Physics Letters

Yumeng Yang